\documentclass[12pt]{iopart}
\usepackage{graphicx}

\begin{document}
\title[Perspectives to Study a Solar CNO Cycle]{Perspectives to Study a Solar CNO Cycle by
Means of a Lithium Detector of Neutrinos.}

\author{A.Kopylov, V.Petukhov}

\address{Institute of Nuclear Research of Russian Academy of Sciences \\
117312 Moscow, Prospect of 60th Anniversary of October Revolution
7A, Russia}

\ead{beril@al20.inr.troitsk.ru}

\begin{abstract}
Lithium detectors have a high sensitivity to CNO neutrinos from
the Sun. The present experimental data and prospects for future experiments on
the detection of CNO neutrinos are discussed. A nonstationary case is considered
when the flux of $^{13}$N neutrinos is higher than the standard solar model predicts
due to the influx of fresh material from the peripheral layers to the solar core.
\end{abstract}
\noindent{\it neutrino experiments, solar abundance, solar interior} 

\maketitle

\section{Introduction}
According to the currently preferred solar models based on the
standard theory of stellar evolution the main source of solar energy
is a pp chain of reactions. The contribution of a CNO cycle to the
energy generated in the Sun is estimated to be 0.8{\%} \cite{1} if
we accept a mass fraction of heavy elements to hydrogen Z/X = 0.0231
$\pm $ 0.0018 according to Grevesse{\&}Sauval (GS98) \cite{2} and
0.5{\%} if we take a recommended value Z/X = 0.0165 $\pm $ 0.0011 of
Asplund, Grevesse{\&}Sauval (AGS05) \cite{3, 4}. Asplund {\it et al.}
applied a time-dependent, 3D hydrodynamical model of the solar
atmosphere instead of 1D hydrostatic model and they have shown that
they have better fits to the Fe lines than conventional 1D
calculations; moreover, they obtained that different lines give
similar abundances. Thus it proved to be a more progressive
technique. However, it has been shown to be in a serious conflict
with the results of helioseismology \cite{5} while GS98 has better
agreed with these data. It was argued \cite{6, 7} and references
therein that there are some indications that a standard solar model
(SSM) needs further improvement, probably to introduce rotation in
the solar core, to get a better agreement with observational data
gained by helioseismology. This conflict has been analyzed in
details by Basu and Antia in \cite{8}. The general conclusion
was that ``the discrepancy caused by revision of solar heavy element
abundances will lead to further improvements in models of the solar
atmosphere and perhaps of the solar interior as well''. Thus the
contradiction is not necessarily an indication that something is
wrong with helioseismology or with the spectroscopic determinations
of the solar photospheric composition. It may just denote that further
corrections to the SSM should be implemented. Some ideas on solar
evolution still are under development, like possible rotation of the
core, rotationally (or gravitationally) induced instabilities or
possible accretion and mass-loss at some stage of solar evolution
etc \cite{7} -- \cite{10}. Although the analyses of some nonstandard
'contributions' may find them to be superfluous, one cannot envisage
everything. It looks very attractive by means of a totally independent
experiment to exclude {\it en masse} some alternatives independent
of their specific nature. For example, it would be very interesting
to get more precise data about the thermonuclear reactions deep in
the core of the Sun complimentary to the results of helioseismology.
In this paper we address the question of what in particular can be
gained from the study of the CNO cycle in the Sun and the prospects
for future experiments. The CNO cycle is the main source of energy for
main sequence stars with a mass and temperature higher than that of
the Sun. Thus it concerns very fundamental questions of stellar evolution.

Figure 1 shows the CNO cycle proposed by Hans Bethe \cite{11} by
which $^{12}$C is converted by protons to $^{14}$N and then back
to $^{12}$C.

\begin{figure}[!ht]
\centering
\includegraphics[width=4in]{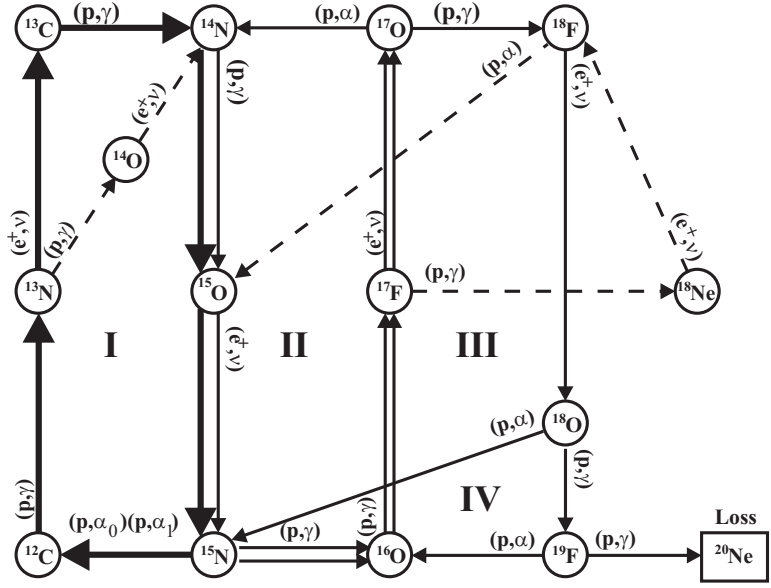}
\caption{The CNO cycle}
\end{figure}

Recently it has been shown in \cite{12} that the accumulation
channel to $^{20}$Ne can be neglected due to very low cross
section of the reaction $^{19}$F + p $\to \quad ^{20}$Ne + $\gamma
$ so that CNO cycle is completely closed at stellar temperatures.
The main body of the cycle is governed by 4 reactions with
hydrogen:

\begin{equation}
\label{eq1}
\left\{ {\begin{array}{l}
 { }^{12}C+p\to { }^{13}N+\gamma \\
 { }^{13}C+p\to { }^{14}N+\gamma \\
 { }^{14}N+p\to { }^{15}O+\gamma \\
 { }^{15}N+p\to { }^{12}C+\alpha \\
 \end{array}} \right.
\end{equation}
\noindent and two $\beta $-decay reactions by which neutrinos are produced:

\begin{equation}
\label{eq2}
\left\{ {\begin{array}{l}
 { }^{13}N\to { }^{13}C+e^++\nu _e \\
 { }^{15}O\to { }^{15}N+e^++\nu _e \\
 \end{array}} \right.
\end{equation}

\noindent Because the half-lives of the isotopes $^{13}$N and $^{15}$O are
very short (9.96 m and 122 s) the fluxes of neutrinos generated in
reactions (\ref{eq2}) are directly connected with the abundances of
isotopes $^{12}$C and $^{14}$N in the solar core where they are
produced. In the SSM it is accepted that primordial
abundances of C and N in the Sun are equal to those currently observed
in the photosphere\footnote[1]{This is not absolutely true because
the effects of gravitational settling and diffusion due to
composition and temperature gradients in SSM lead to decrease of
abundances of metals in the photosphere \cite{14} at a level of
about 10 - 15\%}.

As time has passed $^{12}$C has burned out and $^{14}$N has accumulated in the core of
the Sun. Measurement of neutrino fluxes from the CNO cycle would enable us to find
the primordial abundance of light metals in the Sun, as proposed in \cite{15, 16}. Here we
draw attention to the point that the abundance of $^{12}$C in the core of the Sun is two
orders of magnitude lower than in the peripheral layers, so that even a tiny influx of
fresh material from outside is capable of potentially increasing it substantially while the
abundances of other elements will not be noticeably changed. The result will be that
the flux of neutrinos from the decay of $^{13}$N (here and afterward denoted by f$_{13}$) will be
substantially increased while the flux of neutrinos from the decay of $^{15}$O (let us denote
it f$_{15}$) and of other neutrinos from the pp chain of reactions will not be changed at all.
This is a clear signature of the mass transport between central and peripheral layers. The
question is: can this be accomplished while not changing noticeably the observables fixed
by helioseismology? Of these the most precisely determined are: the mean molecular
weight {$\mu $} in the nuclear active core of the Sun expressed by

\begin{equation}
\label{eq3}
\mu =\frac{4}{6X+Y+2}
\end{equation}

\noindent and the helium abundance in the convection zone. We calculate only the corrections,
not the values themselves which are found in the SSM using the global parameters of
the Sun like mass, radius, luminosity and age, well determined from observations. The
procedure we follow is just built in the SSM with the aim of finding the magnitude of the
corrections and to see what are the prospects for verification by experiment. For example,
in the SSM the diffusion coefficient is not calculated from first principles and usually
includes gravitational settling, diffusion due to composition and temperature gradients.
The mass transport we introduce here is something extra, which we do not specify in
terms "why and how"; it principally differs from the diffusion used in the SSM by not
being spherically symmetric. The masses coming to the core and leaving the core go by
different paths while diffusion is a spherically symmetric process. So when we consider the
mass transport we apply a 3D solar model; this is the most prominent difference from the
SSM, which is a 1D model. By omitting coordinates we neglect the specific character of
the mass transport. Limiting ourselves to the pure evolutionary part of the equations we
find only some averaged values as the estimates for mass transport. We focus here on the
need to test this case experimentally. If the experiment does not observe the anomaly in
f$_{13}$ then all cases independent of their specific nature will be excluded. If the experiment
finds some excess of f$_{13}$, then it will be appropriate to study how exactly a 3D model
should be designed to produce this effect. If there is any excess of f$_{13}$ or not, this is a
question which can only be solved experimentally.

\section{CNO cycle with a mass transport between center and periphery}

Here we address the question of what fluxes of CNO neutrinos are expected in future
experiments if we include in the model some mass transport between the center and the
periphery. The possibility, at least concerning $^{13}$N neutrinos, of obtaining a sufficient
physical effect with very mild mass transport exists due to the big difference in the
abundances of $^{12}$C in the central core and in the outer layers of the radiative zone of
the Sun, so that even a small influx of the fresh material from outside can substantially
change the abundance in the center and, consequently, can increase f$_{13}$. The subtlety in
the interpretation of the experimental data obtained by means of a radiochemical detector
is in the inherently integral result of any radiochemical experiment, i.e. it is not possible
to find the contribution of each neutrino source, only the total rate of production from all
sources. For a lithium detector the important point is how the signal varies in the case
of some mixing in the Sun. The flux f$_{15}$ should decrease while f$_{13}$ should increase in this
case. What will be the net result? To illustrate this let us do a simple exercise. Let us
imagine that mixing in the core occurred at some moment in the past. Here by "moment"
we mean a time interval that is small in comparison with the age of the Sun, say, a few
hundred thousand of years. The neutrino fluxes f$_{13}$ and f$_{15}$ can be found in this case by
solving the set of four differential equations (\ref{eq3}) containing only the nuclear part of the
main loop of the CNO cycle. The idea is to see what will be the production rate in the
lithium detector:

\begin{equation}
\label{eq4}
\left\{ {\begin{array}{l}
 dX(^{12}C)/dt=-\lambda (^{12}C) X(^{12}C)+\frac{12}{15} \lambda (^{15}N) X(^{15}N) \\
 dX(^{13}C)/dt=-\lambda (^{13}C) X(^{13}C)+\frac{13}{12} \lambda (^{12}C) X(^{12}C) \\
 dX(^{14}N)/dt=-\lambda (^{14}N) X(^{14}N)+\frac{14}{13} \lambda (^{13}C) X(^{13}C) \\
 dX(^{15}N)/dt=-\lambda (^{15}N) X(^{15}N)+\frac{15}{14} \lambda (^{14}N) X(^{14}N) \\
 \end{array}} \right.
\end{equation}

The set of equations (\ref{eq4}) has been solved for different zones of the core of the Sun,
starting from the center to the extremity of the core in the approximation that the
temperature profile can be taken as described by the SSM \cite{17} for the present time of
solar evolution. Here it is worth noting that the values $\lambda$ are proportional to the product
$\rho $X($^{1}$H) which can be taken constant in a good approximation during the whole evolution
of the Sun. Figure 2 shows the time evolution for the fluxes f$_{13}$ and f$_{15}$ and also for the
production rate of $^{7}$Be in lithium by CNO neutrinos.

\begin{figure}[!ht]
\centering
\includegraphics[width=4in]{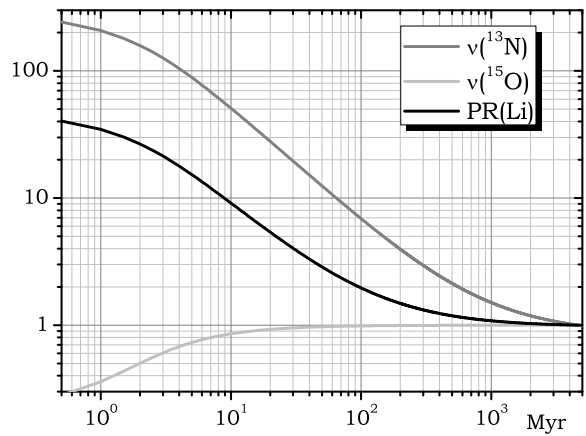}
\caption{The fluxes f$_{13}$ and f$_{15}$ and the production rate
from CNO neutrinos in a lithium detector as a function of time
passed since instantaneous mixing. The fluxes are normalized at present moment of solar evolution.}
\end{figure}

One can see that the flux f$_{15}$ restores the standard level calculated by the SSM
(BS05(OP) \cite{1}) in a time of about 10$^{8}$ years while the flux f$_{13}$ continues to decrease for
a few billion year. As it will be shown later, the real picture is a bit more complicated;
here we use a simplified one to illustrate the interrelations between f$_{13}$ and f$_{15}$ and how
they influence the production rate in lithium. As one can see from Figure 2 the reduced
flux f$_{15}$ is always accompanied by a dramatic increase of the flux f$_{13}$. A very important
and encouraging result for a lithium detector, as one can see from this figure, is that the
production rate from CNO neutrinos is higher in this case than the SSM predicts. The
increase of the effect from $^{13}$N neutrinos surpasses the loss for $^{15}$O neutrinos. This result
will be true independently of whether we use a 1D solar model (SSM) or a 3D solar model
and is explained by the fact that the difference in the abundances of $^{12}$C in the center and
periphery is an order of magnitude higher than of $^{14}$N.

Certainly, care should be taken in performing some manipulations with the standard
model to ensure that it will not distort the observables over the regions allowed by
experiments. Let us see what can be suggested as a possibility not contradicting the
available data. As we mentioned earlier, the mixing in this case should be very mild so
that only the abundance of $^{12}$C in the core of the Sun is increased substantially. One of the
possible ways to do so is the following. We can introduce mixing only in the final phase
of the solar evolution, for example during the last half billion years with a mass transport
coefficient k = 10$^{-10}$ yr$^{-1}$. This means that during that last period of solar evolution the
fresh material with a mass 0.05 of a nuclear active core has been brought from outside
(and has been brought back from the solar core to periphery). We can consider only this
one particular case ("if you see this one you see them all"). If we show that it does not
contradict observations then all other possible cases will be just time variations of this
one, including also the periodic processes. The corresponding set of differential equations
in this case with the initial conditions given by the SSM at the age of 4.1 billion years is
the following:

\begin{equation}
\label{eq5} \left\{ {\begin{array}{l}
 dX(^{12}C)/dt=-\lambda (^{12}C) X(^{12}C)+\frac{12}{15} \lambda (^{15}N) X(^{15}N)+\\\qquad+k
(X_0 (^{12}C)-X(^{12}C)) \\
 dX(^{13}C)/dt=-\lambda (^{13}C) X(^{13}C)+\frac{13}{12} \lambda (^{12}C) X(^{12}C)-k X(^{13}C) \\
 dX(^{14}N)/dt=-\lambda (^{14}N) X(^{14}N)+\frac{14}{13} \lambda (^{13}C) X(^{13}C)+\\\qquad+k
(X_0 (^{14}N)-X(^{14}N)) \\
 dX(^{15}N)/dt=-\lambda (^{15}N) X(^{15}N)+\frac{15}{14} \lambda (^{14}N) X(^{14}N)-k X(^{15}N) \\
 \end{array}} \right.
\end{equation}

Let us recall that the mass transport introduced here by the
coefficient k is a not a spherically symmetric process which makes it very
different from gravitational settling and from the diffusion due to
composition and temperature gradients used by the SSM. Thus we go
beyond the 1D solar model. We do not make an attempt to uncover the
origin of this mass transport or why it took 4.1 billion years for
this process; we just use it to show what processes can be triggered
on a long timescale by instabilities in solar plasma at high
pressure and temperature. Figure 3 shows the evolution of f$_{13}$
and f$_{15}$ for the SSM and for the model where this mixing is
introduced. Practically nothing will be changed in this case except
for the increase of the flux f$_{13}$.

\begin{figure}[!ht]
\centering
\includegraphics[width=4in]{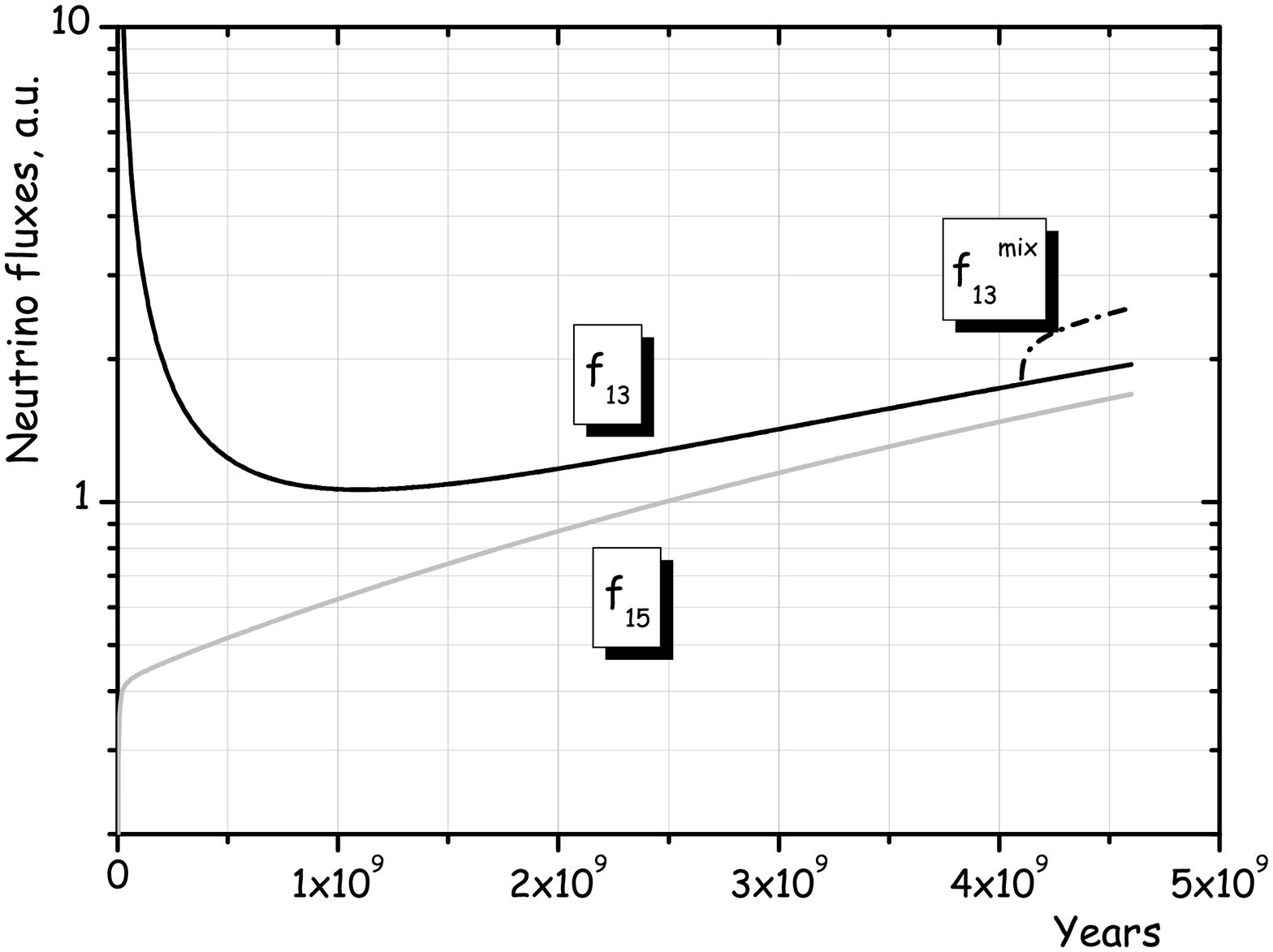}
\caption{The evolution of f$_{13}$ and f$_{15}$ for SSM and for the
model where the mixing is introduced}
\end{figure}

Table 1 contains the corresponding figures for these two cases. The fluxes of f$_{13}$
and f$_{15}$ are presented in relative units. One can see from this table that the substantial
increase (+33.6{\%}) in f$_{13}$ is not accompanied by a noticeable increase in f$_{15}$. The average
molecular weight $\mu $ in the nonstationary case is changed by less than 1{\%}, which does not
seriously contradict the one measured by helioseismology: neither do the changes in the
abundances of $^{12}$C (-2.14{\%}) and $^{14}$N (+7.5{\%}) in the photosphere, which are within the
present uncertainty of about 20{\%}.

\begin{center}
\small
\begin{table}[!h]
\caption{The parameters of the model for standard case and for the
case with mixing.}
\begin{indented}
\item[]\begin{tabular}
{|c|c|c|c|} \hline \hline
& no mix & mix& (mix-no mix)/no mix, {\%} \\
\hline
$X$ & 0,71592 & 0,712992 & -0,40898 \\
$Y$ & 0,26557 & 0,268453 & 1,085589 \\
$ZC$ & $2,99 \cdot 10^{-3}$ & $2,926 \cdot 10^{-3}$ & -2,14047 \\
$ZN$ & $9,27 \cdot 10^{-4}$ & $9,964 \cdot 10^{-4}$ & 7,486516 \\
$f_{13}, present$ & 1,865 & 2,492 & 33,6193 \\
$f_{15}, present$ & 1,594 & 1,592 & -0,12547 \\
$\mu$ & 0,741923 & 0,735008 & -0,93204 \\
\hline \hline
\end{tabular}
\end{indented}
\end{table}
\normalsize
\end{center}

\noindent Here it is worth noting that the change in the abundance ZN is opposite to that
induced by gravitational settling, so that both contributions have a tendency to cancel
each other, while the change in the abundance ZC has the same sign and will be summed
up with the one induced by gravitational settling. So in principle, in future experiments
this mixing can be resolved by this signature from gravitational settling provided the
primordial abundances are known. The abundance of helium in the outer layers of the
Sun has increased by 1{\%} relative to the primordial one, which does not contradict the
current result found from helioseismology Y$_{surf} = 0.248 \cdot ( 1 \pm 0.018)$ \cite{18}.

The possibility for some extra signal in comparison with the predicted one in the
energy range of $^{7}$Be neutrinos principally can be excluded now by the Borexino \cite{19}
experiment. However, it is worth noting that the exact exclusion limit depends critically
upon which model was taken for comparison. As one can see from Table 2 taken from \cite{1},
the flux of $^{7}$Be neutrinos in the BS05 low metallicity model is lower by 10{\%} than in the
BS05 high metallicity model. The current results from Borexino do not help in choosing
the correct model. The 10{\%} decrease of $^{7}$Be neutrinos can be compensated by a 2.1 -- fold
increase of $^{13}$N neutrinos.

\begin{center}
\small
\begin{table}[!ht]
\caption{Predicted solar neutrino fluxes from seven solar models.
The table presents the predicted fluxes, in units of $10^{10}(pp)$,
$10^{9}$($^7$Be), $10^{8}$(pep, $^{13}$N, $^{15}$O), $10^{6}$($^8$B,
$^{17}$F), and $10^{3}$(hep) $cm^{-2}s^{-1}$}
\begin{indented}
\item[]\begin{tabular}
{lcccccccccc} \hline \hline
Model & pp & pep & hep & $^7$Be &  $^8$B & $^{13}$N & $^{15}$O & $^{17}$F \\
\hline
BP04(Yale) & 5.94 & 1.40 & 7.88 & 4.86 & 5.79 & 5.71 & 5.03 & 5.91 \\
BP04(Garching) & 5.94 & 1.41 & 7.88 & 4.84 & 5.74 & 5.70 & 4.98 & 5.87 \\
BS04 & 5.94 & 1.40 & 7.86 & 4.88 & 5.87 & 5.62 & 4.90 & 6.01 \\
BS05($^{14}$N) & 5.99 & 1.42 & 7.91 & 4.89 & 5.83 & 3.11 & 2.38 & 5.97 \\
BS05(OP)& 5.99 & 1.42 & 7.93 & 4.84 & 5.69 & 3.07 & 2.33 & 5.84 \\
BS05(AGS,OP) & 6.06 & 1.45 & 8.25 & 4.34 & 4.51 & 2.01 & 1.45 & 3.25 \\
BS05(AGS,OPAL) & 6.05 & 1.45 & 8.23 & 4.38 & 4.59 & 2.03 & 1.47 & 3.31 \\
\hline
\end{tabular}
\end{indented}
\end{table}
\normalsize
\end{center}

\noindent So we can interpret the results of the Borexino experiment as a confirmation of a
standard solar model with a high Z abundance of GS98 or as a model with a lower Z
abundance according to AGS05 and with the increased flux of f$_{13}$. It means that this is
still an open question and a task for future experiments. Any excess of the signal in the
energy range associated with beryllium neutrinos can be interpreted as a manifestation of
the primordial higher abundance of carbon or as mass transport in the Sun (or both). In
the first case there should also be an increased flux of $^{15}$O neutrinos. This case is especially
attractive for a lithium detector because it has high sensitivity to $^{15}$O neutrinos. In the
second case the ratio $y = f_{13}/f_{15}$ should be higher than expected; let us note also that the
nuclear uncertainties and the ones coming from the neutrino oscillations are canceled in
this ratio, first because of the closure of the CNO cycle and second because the attenuation
factors for $^{13}$N and $^{15}$O neutrinos are very close. From the experimental point of view
the task of measuring the $y = f_{13}/f_{15}$ ratio is very difficult to realize because of the pileup
from different neutrino sources. The problem is really severe: we have four neutrino
sources with intermediate energies $0.5MeV < E_{\nu} < 2.0 MeV$: pep, $^{7}$Be, $^{13}$N, $^{15}$O as one
can see from Figure 4 (adapted from Figure 2 of \cite{1}). Two of them are continuous ($^{13}$N,
$^{15}$O) and two are line sources (pep, $^{7}$Be). The overlap depends upon the type of detector,
its energy resolution etc. In any case, to resolve these neutrino sources the detector should
have an energy resolution comparable with that of semiconductor or cryogenic detectors.
Apparently the present time is not yet ripe for these ideas. Thus the only possible solution
at present seems to be to utilize the different kind of detectors complementarily. Electronic
scintillation detectors and radiochemical detectors appear to be a good match for this
study. The first type gives the differential information on the energy spectra and the
second type the integral information. The electronic detector meets serious problems with
neutrinos from the CNO cycle, partly because of the overlap with $^{7}$Be- and pep-neutrinos
as one can see on Figure 5, and partly because of the background. The radiochemical
lithium detector has comparable sensitivity to all neutrino sources from medium energy
range so its results can be very informative. It always seems very attractive to have the
results obtained by different techniques to achieve higher confidence.

\begin{figure}[!h]
\centering
\includegraphics[width=2.5in, angle=-90]{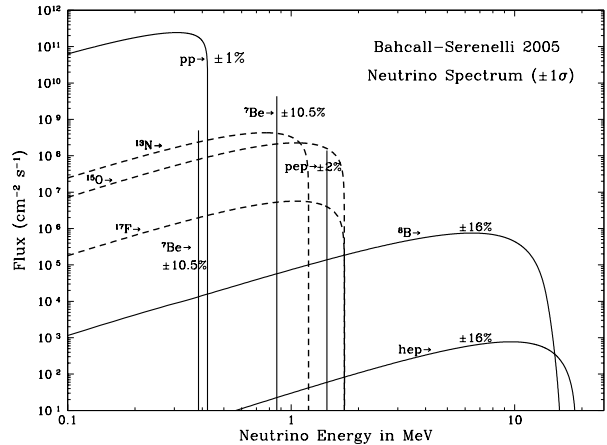}
\caption{Solar neutrino energy spectrum}
\end{figure}

\section{Conclusion. What a lithium experiment can add to our
present knowledge?}

The advantage of a lithium detector is high sensitivity to $^{13}$N
and $^{15}$O neutrinos from the CNO cycle in the interior of the
Sun. Chlorine detector is mainly sensitive to boron neutrinos and
gallium one to pp-neutrinos. Borexino has a very high signal from
$^{7}$Be neutrinos and also from pep-neutrinos at higher
energies. New results from the Borexino experiment \cite{19}
determined the flux of $^{7}$Be neutrinos to be 1.02 of the SSM with
an accuracy 10{\%} under the assumption of the constraint from the
high metallicity SSM and $f_{CNO} < 6.27$ of the SSM (90{\%} CL). Of
course, this is a big achievement; we've made substantial progress
in the accuracy of the determination of the flux of pp-neutrinos
(see Figure 4 of Ref.\cite{19}). But still one should agree that if
one takes the low metallicity SSM the figures will be different.
Because of the overlap of the different neutrino sources sensitivity
to $^{13}$N and $^{15}$O neutrinos is very limited. One can see this
on Figure 5 where the energy spectrum of $\nu $e$^{- }$ scattering
is presented for the ideal resolution.

\begin{figure}[!ht]
\centering
\includegraphics[width=3.5in]{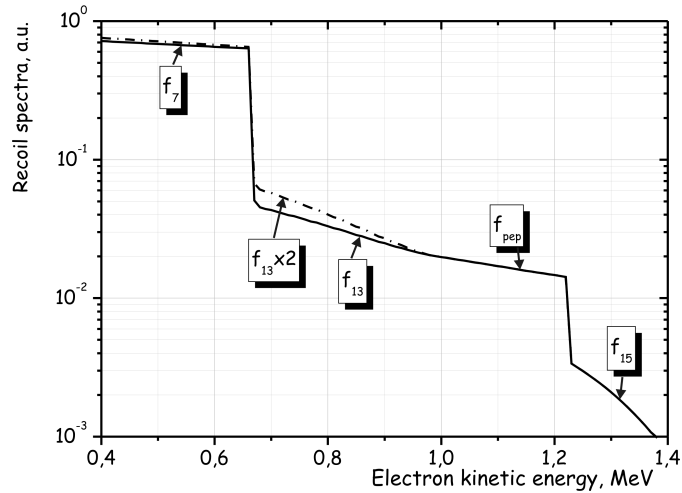}
\caption{The differential energy spectra for recoil electrons for
SSM and for the model with the increased flux f$_{13}$}
\end{figure}

In a lithium detector the contributions of $^{7}$Be, pep and CNO
neutrinos are comparable so it will be very helpful in solving the
important controversy of the low metallicity SSM and helioseismology
and may provide the most valuable piece of information to help
decide which model -- high or low metallicity -- is valid. In the end one
never knows what surprises a new experiment can bring. A chlorine
experiment from the very beginning did not promise a big discovery;
it was just a test of a thermonuclear nature of the generation of
solar energy, and indeed nobody expressed doubts about it. Who could
have thought that it would discover neutrino oscillations? Sometimes
the deviations from the standard behavior are really very tiny, but
the physics behind these tiny deviations is sometimes great. It is
always useful to perform scrupulous measurements to see the real
effect.

In Table 3 the contributions of different neutrino sources for a
lithium detector in comparison with chlorine and gallium detectors
are shown.

\begin{center}
\small
\begin{table}[!h]
\caption{Standard Model Predictions (BP2000): solar neutrino fluxes
and neutrino capture rates without neutrino oscillations, with
1$\sigma$ uncertainties from all sources (combined quadratically)
\cite{21} with new cross-sections measured by LUNA \cite{22}}
\begin{indented}
\item[]\begin{tabular}
{|c|c|c|c|c|} \hline Source& Flux \par (10$^{10}$cm$^{-2}$s$^{-1}$)&
Cl
\par (SNU)& Ga \par (SNU)& Li \par (SNU) \\ \hline pp&
5.99(1.00$^{+0.01}_{-0.01}$)& 0.0& 70.1& 0.0
\\ \hline pep& 1.42$\times $10$^{-2}$(1.00$^{+0.015}_{-0.015}$)& 0.22& 2.8& 9.3
\\ \hline hep& 7.93$\times $10$^{-7}$& 0.04& 0.1& 0.1
\\ \hline $^{7}$Be& 4.84$\times $10$^{-1}$(1.00$^{+0.10}_{-0.10}$)& 1.17& 34.7& 9.2
\\ \hline $^{8}$B& 5.69$\times $10$^{-4}$(1.00$^{+0.20}_{-0.16}$)& 6.49& 13.6& 22.2
\\ \hline $^{13}$N& 3.05$\times $10$^{-2}$(1.00$^{+0.31}_{-0.28}$)& 0.05& 1.9& 1.3
\\ \hline $^{15}$O& 2.31$\times $10$^{-2}$(1.00$^{+0.33}_{-0.29}$)& 0.16& 2.6& 5.7
\\ \hline $^{17}$F& 5.63$\times $10$^{-4}$(1.00$^{+0.25}_{-0.25}$)& 0.0& 0.1& 0.1
\\ \hline Total& & 8.1$^{+1.3}_{-1.1}$& 126$^{+9}_{-7}$& 47.0$^{+6.5}_{-6.0}$
\\ \hline Experiment& &  2.56$^{+0.23}_{-0.23}$&
67.7$^{+3.6}_{-3.6}$&
\\ \hline
\end{tabular}
\end{indented}
\end{table}
\normalsize
\end{center}

\noindent One should take into consideration that the attenuation factors for
boron neutrinos is approximately 0.32 and for all neutrinos of
medium energies it can be taken as 0.56. Taking this into account
one can see that the contribution of CNO neutrinos in a lithium
detector is approximately 18{\%} while the contribution of CNO cycle
to the solar luminosity is only 0.8{\%}.

In the equation of the balance of luminosity of the Sun and of solar
neutrinos the neutrinos from the CNO cycle terminate the equation,
it is the last stroke which fills the gap and which will finally
determine the flux of pp-neutrinos with an accuracy of better than
1{\%}. This is well illustrated by figure 4 of \cite{19} for the
results of Borexino experiment. The measurement of the fluxes of
$^{13}$N and $^{15}$O neutrinos allows the present and the past to
be compared: the present is determined by neutrino fluxes as a probe
of the nuclear activity of the solar core, and the past by the
luminosity of the Sun as its activity is delayed by millions of
years. To know how the Sun shines we should carry out full
spectroscopy of the solar neutrinos as was suggested by Kuzmin,
Zatsepin and Bahcall at the start of solar neutrino research.

\section{Acknowledgements}

The authors thank the referee for comments on the first version of this
paper. It was reported on International Milos (Greece) Symposium ``Physics
of Massive Neutrinos'' 19-23 May 2008 and has been partially supported by
RFBR grant {\#}07-02-00136A and by a grant of Leading Scientific Schools of
Russia {\#}959.2008.2.

\end{document}